\documentclass{PoS}

\usepackage{amsmath}
\usepackage{amsfonts}
\usepackage{amssymb}
\usepackage{epsfig,graphicx,bm}
\usepackage{booktabs}

\title{Quantum groups and braid groups as fundamental symmetries}

\ShortTitle{Quantum groups and braid groups}

\author{\speaker{Niels G. Gresnigt}\thanks{This work is supported in part by XJTLU research grant RDF-14-03-13 and Natural Science Foundation of China grant SCRR0116.}\\
        Xi'an Jiaotong-Liverpool University\\
        E-mail: \email{niels.gresnigt@xjtlu.edu.cn}}

\abstract{The role of quantum groups and braid groups in the description of Standard Model particles is discussed. Some recent results on the use of the quantum group $SU_q(3)$ as a flavour symmetry are reviewed and a connection between two descriptions of Standard Model symmetries, one based on the normed division algebras and the other describing elementary matter as braided objects, is presented.}

\FullConference{EPS-HEP 2017, European Physical Society conference on High Energy Physics\\
		5-12 July 2017\\
		Venice, Italy}

\begin{document}

\section{Introduction}

Quantum groups (which are algebras rather than groups) provide a generalization of the familiar symmetry concepts encoded in Lie groups. Such quantum groups are deformations on Hopf algebras and depend on a deformation parameter $q$ with the value $q=1$ returning the undeformed universal enveloping algebra. First formalized by Jimbo \cite{jimbo1985aq} and Drinfeld \cite{drinfeld1985soviet} as a class of Hopf algebras, quantum groups have found many applications in theoretical physics, see \cite{finkelstein2001q,Finkelstein2012,Steinacker1998,Sternheimer2007,Majid1994,Lukierski2003,castellani1996quantum} and references therein. Likewise, the (Artin) braid groups are an extension of the symmetric groups in which the square of group generators need no longer be equal to the identity. A braid may be thoughts of as an intertwining of strands between two rows of $n$ points, one set above the other.

In 2005, Bilson-Thompson \cite{Bilson-Thompson2005} proposed the Helon model in which the Standard Model (SM) elementary particles are identified as braidings of three ribbons with two crossings. With the additional structure that each ribbon can be twisted (interpreted physically as electric charge) these braids, satisfying certain conditions, map precisely to the first generation fermions of the SM. The original model has since been expanded into a complete scheme for the identification of the SM fermions and weak vector bosons for an unlimited series of generations \cite{Bilson-Thompson2009,Bilson-Thompson2008,Bilson-Thompson2012}. This topological model fits naturally into the context of Loop Quantum Gravity (LQG) which uses spin network graphs with edges labelled by representations of $SU(2)$. Instead labelling the edges by representations of the quantum group $SU_q(2)$ introduces a cosmological constant and requires that the edges be thickened to ribbons \cite{Bilson-Thompson2007}.

Kauffman and Lomonaco have demonstrated, in what they call the Clifford Braiding Theorem, that particular elements of Clifford algebras form representations of (circular) braid groups \cite{kauffman2016braiding}. Clifford (or geometric) algebras have found many applications in physics \cite{hestenes1966sta,doran2003gap,gresnigt2008,gresnigt2007sph} and they are also closely related to the four normed division algebras (NDAs), the reals $\mathbb{R}$, complex numbers $\mathbb{C}$, (non-commutative) quaternions $\mathbb{H}$, and (non-associative) octonions $\mathbb{O}$. One may expect therefore that the NDAs to admit representations of particular braid groups. This is interesting because in 2015 Furey showed that most spacetime and internal symmetries of the SM may be derived starting from only the NDAs acting on themselves \cite{furey2016standard}. In particular, generalized ideals of the complex quaternions give all the SM representations of the Lorentz group whereas those of the complex octonions mirror the behaviour of a single generation of quarks and leptons under unbroken $SU(3)_c$ and $U(1)_{em}$.

The main idea presented here is that using the Clifford Braiding Theorem of Kauffman and Lomonaco it is possible to investigate the braid content of the description of SM symmetries based on the NDAs and compare it to the braided structures found in the Helon model. Remarkably, we find that the braid groups representations admitted by the complex numbers and quaternions are precisely those braid groups present in the Helon model. Further work to investigate the connections between the two models, in particular the role of the octonions and the braid group they represent, is ongoing.

%should  be possible to establish a connection between the topological Helon model of Bilson-Thompson and the approach based on NDAs of Furey. Some evidence to support this hypothesis is given although the details are still currently being investigated by the author.

In section \ref{quantumgroup} we review a recent attempt at taking a quantum group as the broken flavour symmetry for hadrons. Following a review of the Helon model in section \ref{helonmodel} and the NDAs in section \ref{nda}, we then use the Clifford braiding theorem to find the braid group representation admitted by the normed division algebras in section \ref{cbt}. In section \ref{connection} we demonstrate that the braid groups represented by the NDAs are precisely those that appear in the Helon model.
 
%%%%%%%%%%%%%%%%%%%%%%%%%%%%%%%%%%%%%%%%%%%%%%%%%%%%%%%%%%%%%%%%%%%%%%%%%%%%%%%%
\section{Quantum group $SU_q(3)$ as a flavour symmetry}\label{quantumgroup}

In this section we review recent attempts at taking a quantum group as a fundamental flavour symmetry. The use of the classical $SU(3)$ group as a flavour symmetry gives rise to mass relations between the multiplets of the octet and decuplet baryons \cite{gell1961eightfold,okubo1962note}. The Gell-mann-Okubo formula for octet baryons and the Okubo formula for decuplet baryons are
\begin{eqnarray}\label{GMOoctet}
N+\Xi=\frac{3}{2}\Lambda+\frac{1}{2}\Sigma,%\qquad \Omega-\Delta=3(\Xi^*-\Sigma^*).
\end{eqnarray}
and
\begin{eqnarray}\label{GMOdecuplet}
\Omega-\Delta=3(\Xi^*-\Sigma^*),
\end{eqnarray}
respectively.

Gavrilik considered the use of the quantum group $SU_q(3)$ as a flavour symmetry to obtain generalized baryon mass formulas \cite{gavrilik2004quantum,gavrilik2001quantum,gavrilik1998quantum,gavrilik1997quantum}. For suitable values of the deformation parameter $q$ (taken to be a root of unity), the obtained mass formulas are accurate enough that mass differences within baryon multiples, due to electromagnetic mass contributions, can no longer be ignored. Accounting for these electromagnetic mass contribution by using the general QCD parametrization scheme of Morpurgo  \cite{morpurgo1989field}, Gresnigt \cite{Gresnigt2016baryons} derived the following generalized mass formulas for the octet baryons:
%\end{itemize}
\begin{eqnarray}\label{newoctetbaryongeneral}
p+\frac{2[3]_q}{3[2]_q}\Xi^0+\left( \frac{1}{[2]_q-1}-\frac{2[3]_q}{3[2]_q}\right)\Xi^-=
\frac{[3]_q}{[2]_q}\Lambda-\frac{[3]_q}{3[2]_q}\Sigma^0+\left( \frac{1}{[2]_q-1}-\frac{2[3]_q}{3[2]_q}\right)\Sigma^-+\Sigma^+,
\end{eqnarray}
where
\begin{eqnarray}
[N]_q=\frac{q^N-q^{-N}}{q-q^{-1}},
\end{eqnarray}
is a {\small $q$}-number, and {\small $q$} can in general be real or complex. Equation (\ref{newoctetbaryongeneral}) is only valid for the case where $q=q_n=e^{i\theta_8}=e^{\frac{i\pi}{n}}$ for integer $n$. The best fit to experimental data occurs when $n=7$, in which case equation (\ref{newoctetbaryongeneral}) simplifies to:
\begin{eqnarray}\label{newoctet}
p+\frac{(2\Xi^0+\Xi^-)}{3([2]_{q_7}-1)}=\frac{\Lambda}{[2]_{q_7}-1}+\frac{(\Sigma^- -\Sigma^0)}{3([2]_{q_7}-1)}+\Sigma^+.
\end{eqnarray}
Similarly, the Okubo formula for decuplet baryons becomes:
\begin{eqnarray}\label{newdecuplet}
\Omega^--\Delta^-=([2]_q+1)(\Xi^{*-}-\Sigma^{*-}),
\end{eqnarray}
with the best fit occurring when {\small $q=q_n=e^{i\theta_{10}}=e^{\frac{i\pi}{21}}$}. With the values of $q$ so chosen, the errors of these mass relations are only 0.02\% and 0.08\% respectively; a factor of 20 reduction compared to the standard Gell-mann-Okubo mass formulas. These results are summarised in table 1.

\begin{table}
\begin{tabular}{l l l }
\toprule

\textbf{} & \footnotesize{GMO formula (\ref{GMOoctet})} &\footnotesize{Equation (\ref{newoctet})}\\
\midrule
\footnotesize{LHS (MeV)} & \;2257.2 &  \;\;2582.0 \\
\footnotesize{RHS (Mev)} & \;2270.1 &  \;\;2582.6 \\
\footnotesize{Error (\%)} & \;12.9 &  \;\;0.6 \\
\bottomrule
\end{tabular}
\qquad
\begin{tabular}{l l l }
\toprule
\textbf{} & \footnotesize{Okubo formula (\ref{GMOdecuplet})} & \footnotesize{Equation (\ref{newdecuplet})}\\
\midrule
\footnotesize{LHS (MeV)} & \;\;\;\;440.5 &  \;\;440.5 \\
\footnotesize{RHS (Mev)} & \;\;\;\;446.5 &  \;\;440.2 \\
\footnotesize{Error (\%)} & \;\;\;\;6.0 &  \;\;0.3 \\
\bottomrule
\end{tabular}
\caption{Decuplet baryon mass formula.}
\end{table}

Additionally, the use of a quantum group as a flavour symmetry suggest an explicit formula for the Cabibbo angle, taken to be $\frac{\pi}{14}$, in terms of the deformation parameter $q$ and spin parity $J^P$ of the baryons \cite{gavrilik2004quantum,Gresnigt2016baryons,gavrilik2001can}
\begin{eqnarray}\label{qcabibbo}
\theta_C=-iJ^P\ln{q}.
\end{eqnarray} 

%%%%%%%%%%%%%%%%%%%%%%%%%%%%%%%%%%%%%%%%%%%%%%%%%%%%%%%%%%%%%%%%%%%%%%%%%%%%%%%%%%
\section{The Helon model}\label{helonmodel}

The Helon model of Bilson-Thompson maps the simplest non-trivial braids consisting of three (twisted) ribbons and two crossings to the first generation of SM fermions. There are exactly 15 states in the model with no place for a right-handed neutrino. Quantized electric charges of particles are represented by integral twists of the ribbons of the braids; The color charges of quarks and gluons are accounted for by the permutations of twists on certain braids; and simple topological processes are identified with the electroweak interaction, the color interaction, and conservation laws.  

These braided structures may be embedded within a larger network of braided ribbons. Such a braided ribbon network (BRN) is a generalization of a spin network, fundamental in LQG. The embedding of the braids into BRNs make it possible to 
develop a unified theory of matter and spacetime in which both are emergent from the BRN \cite{Bilson-Thompson2007}. The representation of first generation SM fermions in terms of braids and an illustration of the embedding of such a braid into a BRN are shown in Figure 1. 

\begin{figure}[h!]
\centering
   \includegraphics[scale=0.21]{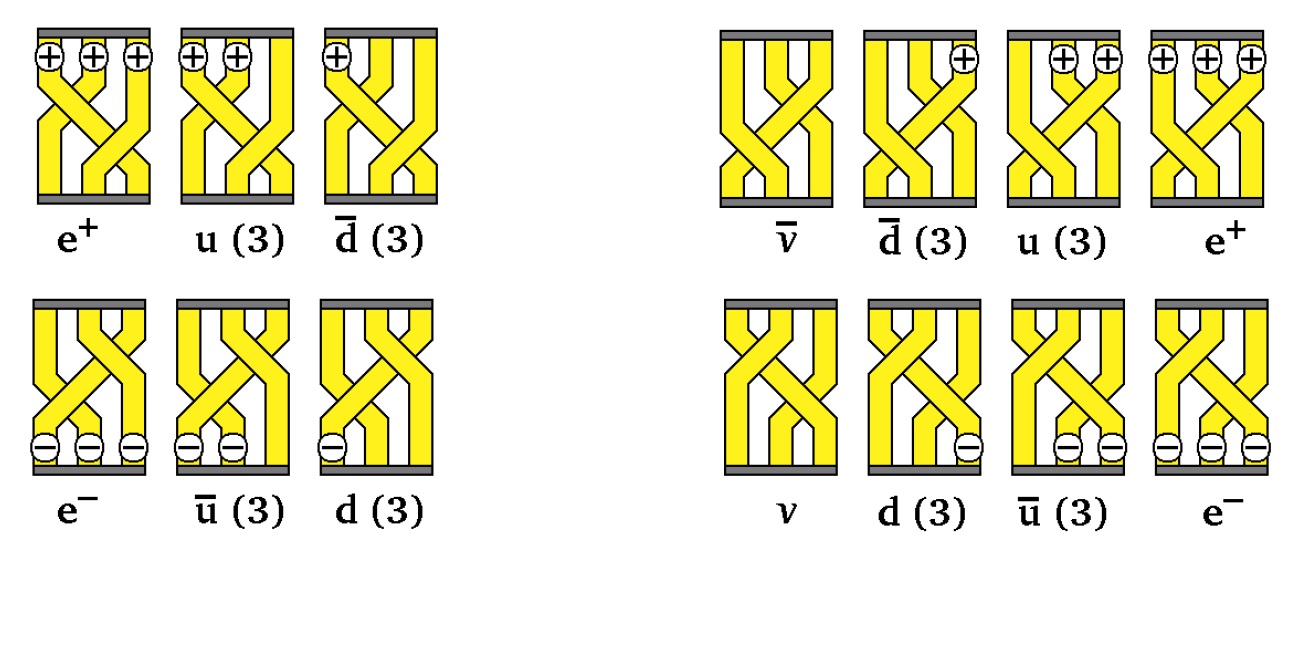}
\includegraphics[scale=0.21]{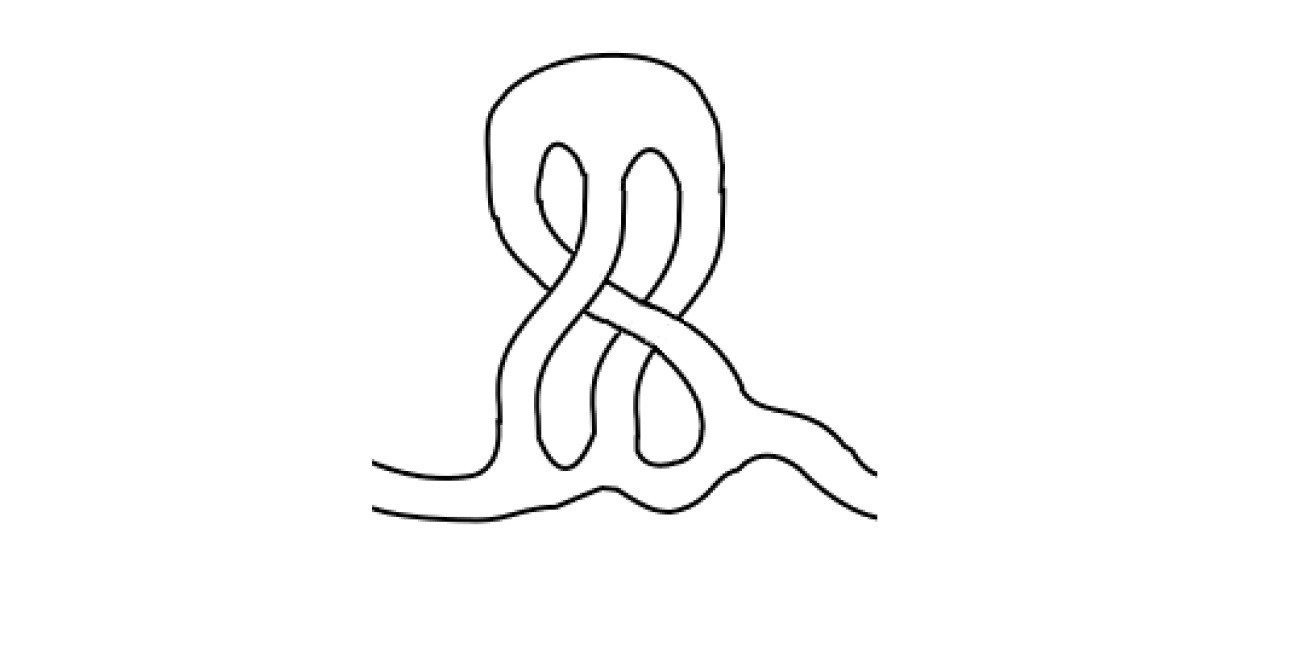}
\caption{The Helon model of Bilson-Thompson in which the first generation SM fermions are represented as braids of three (possibly twisted) ribbons, and the embedding of such braids into a braided ribbon network. Sources, \cite{Bilson-Thompson2005} and \cite{Bilson-Thompson2007}.}
\end{figure}
\vspace{-0.5cm}
%%%%%%%%%%%%%%%%%%%%%%%%%%%%%%%%%%%%%%%%%%%%%%%%%%%%%%%%%%%%%%%%%%%%%%%%%%%%%%%%%
\section{Normed division algebras}\label{nda}

A division algebra is an algebra over a field where division is always possible, with the exception of division by zero. Nature admits only four normed division algebras over the reals, the real numbers $\mathbb{R}$, the complex numbers $\mathbb{C}$, the quaternions $\mathbb{Q}$ and the octonions $\mathbb{O}$. Starting from the real numbers and generalizing to the complex numbers, one has to give up the ordered property of the reals. Generalizing in turn to the quaterions one furthermore gives up the commutatativity of the reals and complex numbers. The quaternions are spanned by ${1,I,J,K}$ with $1$ being the identity and $I,J,K$ satisfying
\begin{eqnarray}
I^2=J^2=K^2=IJK=-1.
\end{eqnarray}

Finally, in moving to the octonions one has to give up the associativity of the reals, complex numbers, and quaternions. The lack of associativity of the octonions means their applications to physics have not been studied in as much details as for the other normed division algebras. An excellent introduction to the octonions is given by Baez \cite{baez2002octonions}. The octonions are spanned by the identity $1=e_0$ and seven $e_i$ satisfying
\begin{eqnarray}
e_ie_j=-\delta_{ij}e_0+\epsilon_{ijk}e_k,
\end{eqnarray}
where 
\begin{eqnarray}
e_ie_0=e_0e_i=e_i,\;e_0^2=e_0,
\end{eqnarray}
and $\epsilon_{ijk}$ is a completely antisymmetric tensor with value +1 when $ijk = 123,\;145,\;176,\;246,\;257,\\347,\;365$. The multiplication of quaternions and octonions is shown in Figure 2.
\begin{figure}
\centering
\includegraphics[width=0.20\linewidth]{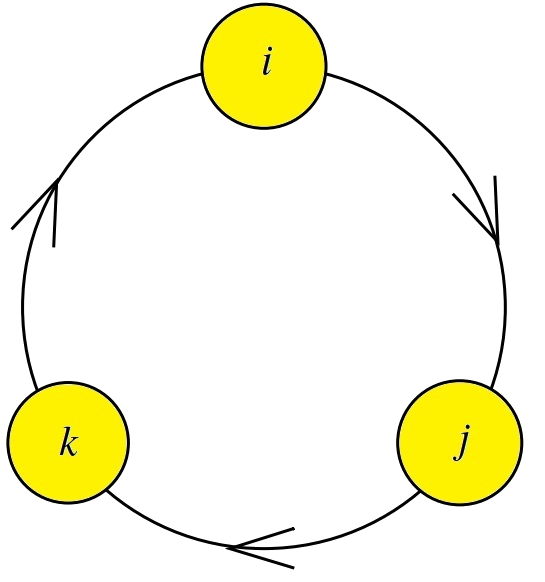}
\qquad\qquad\qquad
\includegraphics[width=0.25\linewidth]{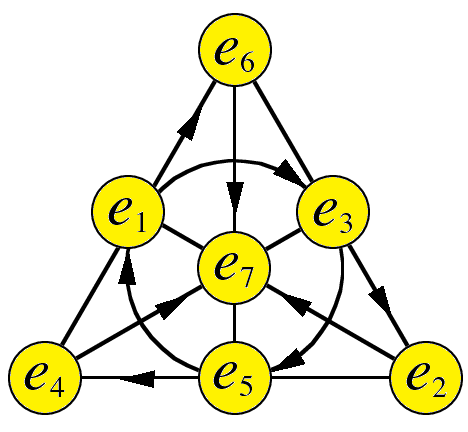}
\caption{Quaternion multiplication $I^2=J^2=K^2=IJK=-1$, and octonion multiplication represented using a Fano plane.}
\end{figure}

Every straight line in the Fano plane of the octonions (together with the identity) generates a copy of the quaternions, for example ${e_4,e_1,e_6}$. The circle ${e_1,e_3,e_5}$ also gives a copy of the quaternions, making for a total of seven copies of the quaternions in the octonions.

It was very recently (2015) shown by Furey that one can represent much of SM physics in terms of the NDAs acting on themselves \cite{furey2016standard}. Although we will not go into the details here, the main results are that the complex quaternions, $\mathbb{C}\otimes \mathbb{H}$ admit all the representations of the Lorentz group that one finds in the SM including the scalar, vector, spinor, and field strength tensor representations, whereas the complex octonions $\mathbb{C}\otimes \mathbb{O}$ provide a representations of the unbroken internal symmetry $SU(3)\times U(1)$ of a single generation of SM fermions. That is, within the Dixon algebra, defined to be the product $\mathbb{R}\otimes\mathbb{C}\otimes\mathbb{H}\otimes\mathbb{O}$, one recovers both the spacetime and internal symmetries of SM matter.

%%%%%%%%%%%%%%%%%%%%%%%%%%%%%%%%%%%%%%%%%%%%%%%%%%%%%%%%%%%%%%%%%%%%%%%%%%%%%%%%%
\section{Clifford algebra representations of circular Artin braid groups}\label{cbt}

The Artin braid group on $n$ strands is denoted by $B_n$ and is generated by elementary braids $\left\lbrace \sigma_1,...,\sigma_{n-1}\right\rbrace$ subject to the relations
\begin{eqnarray}\label{braidrelations}
\sigma_i \sigma_j=\sigma_j\sigma_i,\;\textrm{whenever}\; \vert i-j \vert > 1,\\
\nonumber\sigma_i\sigma_{i+1}\sigma_i=\sigma_{i+1}\sigma_i \sigma_{i+1},\;\textrm{for}\; i=1,....,n-2.
\end{eqnarray}
The braid groups $B_n$ are an extension of the symmetric groups $S_n$ with the condition that the square of each generator being equal to one lifted. It was recently shown by Kauffman and Lomonaco that Clifford algebras admit algebras representations of braid groups \cite{kauffman2016braiding}.

For a Clifford algebra $Cl(n,0)$ over the real numbers generated by linearly independent elements $\left\lbrace e_1,e_2,...,e_n\right\rbrace $ with $e_k^2=1$ for all $k$ and $e_ke_l=-e_le_k$ for $k\neq l$, the algebra elements $\sigma_k=\frac{1}{\sqrt{2}}(1+e_{k+1}e_k)$ form a representation of the circular\footnote{A circular braid on $n$ strings has $n$ strings attached to the outer edges of two circles which lie in parallel planes in $R^3$.} Artin braid group $B_n$. This means that the set of braid generators $\left\lbrace  \sigma_1,\sigma_2,...,\sigma_n\right\rbrace $ where
\begin{eqnarray}
\sigma_k &=& \frac{1}{\sqrt{2}}(1+e_{k+1}e_k),\;\textrm{whenever}\; 1\leq k <n,\\
\sigma_n &=&\frac{1}{\sqrt{2}}(1+e_1e_n),
\end{eqnarray}
satisfy the braid relations (\ref{braidrelations}). Although the original theorem as found in \cite{kauffman2016braiding} assumes that $e_k^2=1$ for all $k$, the proof likewise holds when $e_k^2=-1$, as is easily checked. The important point is that it fails to hold when we have a general Clifford algebra $Cl(+,-)=Cl(p,q)$ of mixed signature.

The braid generators live in the even part of $Cl(n,0)$, denoted by $Cl^{+}(n,0)$. The even part of a Clifford algebra $Cl^{+}(n,0)$ is always isomorphic to the Clifford algebra of one less dimension but with opposite signature, $Cl^{+}(n,0)\cong Cl(0,n-1)$. We will make use of this isomorphism in what follows.

%%%%%%%%%%%%%%%%%%%%%%%%%%%%%%%%%%%%%%%%%%%%%%
\subsection{The complex numbers and a representation of the Artin braid group $B_2$}

The complex numbers $\mathbb{C}$ with basis $\left\lbrace 1, i\right\rbrace$ are isomorphic to the Clifford algebra $Cl(0,1)$ which has $e_1^2=-1$. Given the isomorphism above it means we also have an isomorphism with $Cl^+(2,0)$, the even part of $Cl(2,0)$. 

Therefore, via the Clifford Braiding Theorem, the complex number algebra $\mathbb{C}$ admits a representation of the braid group $B_2$. In this case the Artin braid group is equivalent to the circular Artin braid group $B^c_2\cong B_2$. The single braid generator $\sigma_1$ can be represented in terms of the scalar and bivector of $Cl(2,0)$, so that
\begin{eqnarray}
\sigma_1=\frac{1}{\sqrt{2}}(1+e_2e_1),\qquad \sigma_1^{-1}=\frac{1}{\sqrt{2}}(1-e_2e_1),
\end{eqnarray} 
with the inverse generators defined by inserting a minus sign in front of the bivector terms. Using the isomorphism $Cl^{+}(2,0)\cong Cl(0,1)\cong\mathbb{C}$, $\mathbb{C}$ gives a representation of $B_2$ with the braid generator expressed as
\begin{eqnarray}
\sigma_1=\frac{1}{\sqrt{2}}(1+i),\qquad \sigma_1^{-1}=\frac{1}{\sqrt{2}}(1-i).
\end{eqnarray}

%%%%%%%%%%%%%%%%%%%%%%%%%%%%%%%%%%%%%%%%%%%%%%
\subsection{The quaternions and a representation of the circular Artin braid group $B^c_3$}

Moving on to the quaternions $\mathbb{H}$, one can use the isomorphism $\mathbb{H}\cong Cl(0,2)\cong Cl^+(3,0)$ to find a quaternionic representation of the braid group $B^c_3$. $Cl(0,2)$ is spanned by ${1,e_1,e_2,e_1e_2=e_{12}}$ with
\begin{eqnarray}
e_1^2=e_2^2=e_{12}^2&=&e_1e_2e_{12}=-1,\\
e_1e_2=-e_2e_1,\; e_1e_{12}&=&-e_{12}e_1,\; e_2e_{12}=-e_{12}e_2.
\end{eqnarray}
One can thus identify $e_1=I,\;e_2=J,\;e_{12}=K$ to obtain a copy of $\mathbb{H}$. The even subalgebra of   $Cl(3,0)$ contains six bivectors (and the scalar), three of which may be related to braid generators for $B^c_3$
\begin{eqnarray}
\sigma_1=\frac{1}{\sqrt{2}}(1+e_2e_1),\; \sigma_2=\frac{1}{\sqrt{2}}(1+e_3e_2),\; \sigma_3=\frac{1}{\sqrt{2}}(1+e_1e_3).
\end{eqnarray}
It is readily checked that
\begin{eqnarray}
\sigma_1\sigma_2\sigma_1=\sigma_2\sigma_1\sigma_2,\;\;
\sigma_2\sigma_3\sigma_2=\sigma_3\sigma_2\sigma_3,\;\;
\sigma_3\sigma_1\sigma_3=\sigma_1\sigma_3\sigma_1.
\end{eqnarray}
In terms of the quaternions we have
\begin{eqnarray}
\nonumber\sigma_1&=\frac{1}{\sqrt{2}}(1+I),\;\;\sigma_2=\frac{1}{\sqrt{2}}(1+J),\;\;\sigma_3=\frac{1}{\sqrt{2}}(1+K),
\end{eqnarray}
with the inverses again obtained by inserting a minus sign. A final important point to note is that the order of these braid generators defined in terms of Clifford algebra bivectors is eight. 
%%%%%%%%%%%%%%%%%%%%%%%%%%%%%%%%%%%%%%%%%%%%%%
\subsection{The octonions and a representation of the circular Artin braid group $B^c_7$}

In addition to being non-commutative, the octonion algebra $\mathbb{O}$ is also non-associative making a matrix representation of the algebra impossible. However, by defining a standard ordering to any product of octonions, it is possible to recover an associative description of the octonions. Let $n$, $m$, and $p$ be three octonions. One defines the octonion chain $\overleftarrow{pnm}$ as the map $\overleftarrow{pnm}:f\mapsto p(n(mf))$. One can generalize this to a product of arbitrary many octonions. The resulting algebra is the chained octonions $\overleftarrow{\mathbb{O}}$ which is associative and can be shown to be isomorphic to $Cl(0,6)$ \cite{furey2016standard}. 

Using the isomorphism $\overleftarrow{\mathbb{O}}\cong Cl(0,6)\cong Cl^+(7,0)$ one finds a representation of the braid group $B^c_7$ in terms of the chained octonions
\begin{eqnarray}
\sigma_i&=\frac{1}{\sqrt{2}}(1+\overleftarrow{e_{i+1}e_{i}}),\;\;\sigma_7=\frac{1}{\sqrt{2}}(1+\overleftarrow{e_1e_7}).
\end{eqnarray}

In summary then, we have
\begin{eqnarray}
\nonumber\mathbb{C}\rightarrow B_2:\; &\sigma_1&=\frac{1}{\sqrt{2}}(1+i),\;\;\sigma_1^{-1}=\frac{1}{\sqrt{2}}(1-i),\\
\nonumber\mathbb{H}\rightarrow B^c_3:\; &\sigma_1&=\frac{1}{\sqrt{2}}(1+I),\;\;\sigma_2=\frac{1}{\sqrt{2}}(1+J),\;\;\sigma_3=\frac{1}{\sqrt{2}}(1+K),\\ 
\nonumber\overleftarrow{\mathbb{O}}\rightarrow B^c_7:\; &\sigma_i&=\frac{1}{\sqrt{2}}(1+\overleftarrow{e_{i+1}e_{i}}),\;\;\sigma_7=\frac{1}{\sqrt{2}}(1+\overleftarrow{e_1e_7}).
\end{eqnarray}

%%%%%%%%%%%%%%%%%%%%%%%%%%%%%%%%%%%%%%%%%%%%
\section{Common elements of the Helon model and normed division algebra model}\label{connection}

The main new idea of this paper is that the Clifford Braiding Theorem can be applied to the NDAs to investigate their braid content. The braid group representations of the NDAs presented above make it possible to compare this approach to the Helon model where elementary particles are represented as braided ribbons.

The complex quaternions $\mathbb{C}\otimes\mathbb{H}$ admit all the SM representations of the Lorentz group \cite{furey2016standard}. Given that the complex numbers and quaternions also contain representations of the braid groups $B_2$ and $B_3^c$ respectively, this suggests it should be possible to represent SM fermions as elements of (some product of) the braid groups $B_2$ and $B^c_3$.

Looking now at the Helon model we encounter precisely these two braid groups. The twisting of the ribbons, representing electric charge corresponds to elements of $B_2$ (which is isomorphic to the integers). When the ribbon is twisted the two edges of the ribbon braid one another. Additionally, the braiding of three ribbons forms a braid word in $B_3$. Finally, the embedding of Helon model braids into a BRN requires the braid to be capped, by which we mean that the ribbons are connected at the top and bottom to disks, or framed nodes. This means we have not just braids in $B_3$ but rather braids in the circular braid group $B_3^c$.

Because in general the twisting of the ribbons and the braiding of ribbons is not commutative, with the braidings inducing a permutation on the twists of the ribbons, the mathematical structure is that of the semi-direct product $B_3^c \ltimes (B_2)^3$. Since $B_2 \cong \frac{1}{2}\mathbb{Z}$ we can rewrite this as $B_3^c \ltimes (\frac{1}{2}\mathbb{Z})^3=B_3^c \ltimes (\frac{1}{2}\mathbb{Z}\times \frac{1}{2}\mathbb{Z}\times \frac{1}{2}\mathbb{Z})$. We denote an element of $(\frac{1}{2}\mathbb{Z}\times \frac{1}{2}\mathbb{Z}\times\frac{1}{2}\mathbb{Z})$ by a vector $[a_1,a_2,a_3]$ of multiples of half integers as in \cite{Bilson-Thompson2009,Bilson-Thompson2008}. A general Helon model braid may then be written as $(\Lambda, [a_1,a_2,a_3])$ where $\Lambda\in B^c_3$ is the braid word and $[a_1,a_2,a_3]\in (\frac{1}{2}\mathbb{Z})^3$ is the twist word.

We can multiply two framed braids together by first joining the bottom of the ribbons of the first braid to the tops of the ribbon of the second braid and then sliding (isotop) the twists from each component braid downward. Doing so, the twists carried by the first braid will get permuted by the second braid. We will write the resulting product in standard form with the braiding first followed by the twisting. We can then write the composition law as 
\begin{eqnarray}
&(&\Lambda_1,[a_1,a_2,a_3])(\Lambda_2,[b_1,b_2,b_3])\\
\nonumber&=&(\Lambda_1\Lambda_2, P_{\Lambda_2}([a_1,a_2,a_3])+[b_1,b_2,b_3]),\\
\nonumber&=&(\Lambda_1\Lambda_2, ([a_{\pi(\Lambda_2)(1)},a_{\pi(\Lambda_2)(2)},a_{\pi(\Lambda_2)(3)}])+[b_1,b_2,b_3]),\\
\nonumber&=&(\Lambda_1\Lambda_2, ([a_{\pi(\Lambda_2)(1)}+b_1,a_{\pi(\Lambda_2)(2)}+b_2,a_{\pi(\Lambda_2)(3)}+b_3])
\end{eqnarray}
where $\Lambda_1$ and $\Lambda_2$ are two braid words, $P_{\Lambda_i}$ is the permutation induced on $[a,b,c]$ by the braid word $\Lambda$, and $\pi:B_3^c\rightarrow S_3$ with $\pi(\sigma_1)=(12)$, $\pi(\sigma_2)=(23)$, $\pi(\sigma_3)=(31)$.

%%%%%%%%%%%%%%%%%%%%%%%%%%%%%%%%%%%%%%%%%%%%%%%%%%%%%%%%%%%%%%%%%%%%%5
\section{Discussion}

Quantum groups and braid groups play important roles in physics. Taken as a flavour symmetry, $SU_q(3)$ leads to exceptionally accurate baryon mass formulas. It would be interesting to consider the deformation of other internal symmetries in the SM and extend previous work to calculate cross sections and branching ratios of particles when one assumes underlying quantum group symmetries. Quantum groups have also played a role in LQG where labelling the edges of spin networks by representations of $SU_q(2)$ introduces a (positive) cosmological constant. At the same time, doing so requires replacing the ended of the network by ribbons, giving a braided ribbon network.

The main idea presented here is that the Clifford Braiding Theorem it is possible to investigate the braid content of the description of SM symmetries based on the NDAs and compare it to the braided structures found in the Helon model. Remarkably, we find that the braid groups representations admitted by the complex numbers and quaternions are precisely those braid groups present in the Helon model. Further work to investigate the connections between the two models, in particular the role of the octonions and the braid group they represent, is ongoing.

The choice of braid group structure in the Helon model is somewhat arbitrary, chosen because it is the seemingly simplest possibility. Using NDAs provides a justification for why one should consider braids consisting of three ribbons, rather than any other number of ribbons. That is, the four NDAs dictate which braid groups one should consider. What is still needed is a mechanism to determine which elements within each braid groups should be considered. It may be speculated that this could be determined by the finite order of the braid generators as represented by the NDAs. In these cases the order of each generator is eight, in contrast to the standard braid generators which have infinite order. This is currently work in progress.

%Further work to investigate the connections between the two models, in particular the role of the octonions and the braid group they represent, is ongoing. In the NDA model the complex octonions give rise to the unbroken internal symmetries $U(1)_{em}$ and $SU(3)_c$. The (chained) octonions admit a representation of $B^c_7$. A next step is to study this braid group in the context of the Helon model and see if it is possible to connect it with the internal symmetries. There is some indication that this my be possible. In the Helon model, the color force is represented by stacking three braid on top of one another. Inside the complex octonions one can likewise find multiple copies of the complex quaternions.

%%%%%%%%%%%%%%%%%%%%%%%%%%%%%%%%%%%%%%%%%%%%%%%%%%%%%%%%%%%%%%%%%%%%%5
\footnotesize\section*{Acknowledgments}
\footnotesize The author wishes to thank Cohl Furey, Louis Kauffman, Adam Gillard, and Benjamin Martin for insightful discussions and communications.

%%%%%%%%%%%%%%%%%%%%%%%%%%%%%%%%%%%%%%%%%%%%%%%%%%%%%%%%%%%%%%%%%%%%%
\bibliography{NielsReferences}  % Replace xxx by your  usercode (no extension)
\bibliographystyle{unsrt}  
%%%%%%%%%%%%%%%%%%%%%%%%%%%%%%%%%%%%%%%%%%%%%%%%%%%%%%%%%%%%%%%%%%%%%

%\end{thebibliography}

\end{document}